\documentclass[12pt]{iopart}
\usepackage{iopams}
\jl{6}

\def\sch{Sch\"odinger's }                

\begin{document}
\title{Quantum fields as gravitational sources}
\author{Mark J Hadley}

\address{Department of Physics, University of Warwick, Coventry
CV4~7AL, UK\\ email: Mark.Hadley@warwick.ac.uk}

\begin{abstract}
The practice of setting quantum fields as sources for classical general relativity is examined. Several conceptual problems are identified which invalidate apparently innocuous equations. Alternative ways to links classical general relativity with quantum theory using Bohm's theory are proposed.
\end{abstract}

\submitted
\maketitle

\section{Introduction} \label{introduction}

In a recent paper Carlip \cite{carlip} investigates the implications of combining classical general relativity with quantum theory as an alternative to replacing general relativity with a quantum theory of gravity, which has so far remained elusive. Carlip considers the equation:

\begin{equation}
G_{ab} = 8\pi\hat{T}_{ab}
\end{equation}
and rightly rejects it as equating a c-number with an operator. The author then considers:
\begin{equation}
G_{ab} = 8 \pi<\psi|\hat{T}_{ab}|\psi>
\label{eq:vec}
\end{equation}
and evaluates the experimental consequences. Carlip is by no means the first to use such a semiclassical approximation, and acknowledges some minor issues. Although \ref{eq:vec} may read sensibly mathematically and may seem to be a convenient working approximation in the absence of a full unifying theory, it is untenable as an equation of physics for a variety of reasons that are developed below.

Note that $<\psi|\hat{T}_{ab}|\psi>$ usually denotes the total, integrated, expected values of energy momentum etc, but in General Relativity the RHS is an energy momentum density at a particular spacetime point. To adopt a more conventional notation we should have:

\begin{equation}
G_{ab}(x,t) = 8 \pi\int \psi^*(x',t)\hat{T}_{ab}(x,t)\psi(x',t)dx'
\label{eq:vec3}
\end{equation}
or more simply:
\begin{equation}
G_{ab}(x,t) = 8 \pi\psi^*(x,t)\hat{T}_{ab}\psi(x,t)
\label{eq:vec2}
\end{equation}

It should also be noted that the RHS is of only limited applicability for quantum theory because it applies only to pure states. Unpolarised electron beams, for example, cannot be described using equation~\ref{eq:vec2}. Quantum states form a convex set with pure states on the boundary. A general mixed state cannot be defined as a state vector but only as a non-unique mixture of state vectors (See \cite{geldart} for one of the  clearest explanations). A generalisation of the RHS to include all quantum states is well known, it uses the state operator (density matrix) rather than a state vector:

\begin{equation}
G_{ab}(x,t) = 8 \pi \hat{\rho}(x,t) \hat{T}_{ab}
\label{eq:eq}
\end{equation}

For a pure state \ref{eq:vec2} and \ref{eq:eq} are equivalent, but we shall see that the latter form helps to clarify the conceptual inconsistencies.

The equation is fascinating both because it implies that we can define the gravitational field (space time curvature) due to a single electron and it can also be tested for macroscopic objects because the equations and results of quantum theory are universal and apply equally to a 100kg weight as to an electron. Some conceptual problems might be expected because the LHS is based on a classical theory but the RHS is based on quantum theory that cannot be described by a local hidden variable theory.

\section{Some problems}

The problems with equation~\ref{eq:eq} stem from the fact that it relates the actual curvature of space time to the average of the values of the energy/momentum that would be obtained if a measurement were made. It is a most peculiar equality to make, in my opinion no more sensible than equating a c-number to an operator which was discounted. They are different sorts of object.

Given equation~\ref{eq:eq} there will be instantaneous (non-local) changes to the RHS when a measurement is made. An example would be a an electron with a well-defined momentum, for which the energy-momentum density would be low and spatially extensive. If a position measurement was then made at one place the energy-momentum density would become highly localised. At spacelike separated locations the wavefunction, and hence the energy -momentum density, would drop to zero. If this gravitational field were observable then faster than light communication would be possible. But even if it were not measurable, then $G_{ab}$ would still be subject to discontinuities.

Consider a macroscopic test which uses the fact that the mixed states described by \ref{eq:eq} includes both classical mixed states as well as mixed quantum states. A classical mixed state is where a number of different outcomes are possible, each with its own probability (which add up to one); such a state operator would be diagonal because the off-diagonal elements describe quantum entanglement.

We place a massive, 100kg, ball on top of a hill next to a mechanism controlled by a radioactive source such that if a decay takes place in a specified time interval then the ball rolls to the right, otherwise it rolls to the left, with equal probability. Like \sch cat experiment, the wavefunction is a superposition of states with and without a decay. Because it is a massive body, interaction of with the environment causes rapid decoherence. This can be thought of as the collapse of the wavefunction or mathematically by considering interactions with the environment and then taking the trace to leave a state operator for the ball alone. The state operator is now of diagonal form: 50\% on the left of the hill, 50\% on the right. However the RHS is still a probability function, though now in the usual sense - it says there is a probability of 50\% that the ball is on the left and 50\% that it is on the right - just as we might write for the toss of a coin before we look. However equating the curvature of spacetime to the probability function is now clearly ridiculous. The curvature would be equivalent to a 50Kg ball at each position.

In both the quantum entangled case and the classical mixture the probability function expresses a degree of ignorance. In the classical case the ignorance can be reduced arbitrarily by subsequent measurements. In the quantum case, ignorance of one parameter (eg position) can be replaced by ignorance of a complimentary parameter (eg momentum) by a measurement. In both the classical and quantum case, the expectation values change abruptly and instantaneously and (non-locally) when a measurement is made because information changes. In both cases $G_{ab}$ would be discontinuous.

One possible excuse for working with equation \ref{eq:eq} is that you are interested in a macroscopic limit of large numbers of particles. Just as classical electromagnetic waves can be constructed from a distribution of single photon events. But there is no limiting length scale on the validity of $G_{ab}$. Even for a large number of particles there will still be a length scale at which the wavefunction expression is a probability of finding a single particle in a particular very small region of space time. The only way round that is to limit the equations of relativity in some way so that they only apply over macroscopic regions. e.g.

\begin{equation}
<G_{ab}(x',t)>_{R(x,t)} = 8 \pi <\hat{\rho}(x',t) \hat{T}_{ab}>_{R(x,t)}
\label{eq:eqav}
\end{equation}
where $<...>_{R(x,t)}$ denotes an average over a small region of spacetime centred on $(x,t)$. This is essentially the approach taken in cosmology where the stars and galaxies are approximated as a smooth matter distribution and a smooth metric is sought. However for an equation of fundamental physics or as a union of classical relativity and quantum theory it is surely unsatisfactory to be taking an average of some undefined microstates.

An alternative average would be over an ensemble of similar systems, $E$:
\begin{equation}
<G_{ab}(x,t)>_E = 8 \pi <\hat{\rho}(x,t) \hat{T}_{ab}>_E
\label{eq:eqen}
\end{equation}

But this expression implies the existence of individual $G_{ab}(x,t)$ over which the average is taken. The RHS of equation~\ref{eq:eq} can legitimately be taken as the expectation value of an arbitrary large ensemble, but the 100kg example shows that for finite numbers it can only be an approximation (For odd numbers of balls, there is inevitably a 50Kg discrepancy between the average and the actual mass at a location).

\section{A possible solution}
An alternative resolution is offered by Bohm's theory. Since spacetime curvature in an asymptotically flat region can be used to define position, momentum and all components of spin that implies a hidden variable theorem. It is not important that these quantities can be measured to arbitrary precision to yield useful information, only that they are well defined in the theory. Bohm's theory is the best developed hidden variable theory that is compatible with quantum theory. It ascribes a well defined position and momentum to particles at all times. It is a trajectory theory, where particles follow a trajectory, $x'(t)$ in spacetime in such a way as to give the results of quantum theory. With a wavefunction $\psi = R \exp(-i S/\hbar)$ which satisfies \sch equation, the energy and momentum are calculated as :

\begin{eqnarray}
T_{00}(x,t) &=& m\delta(x-x'(t),t) + Q(x,t) \\
T_{0i} &=& m \nabla_i S(x,t)\delta(x-x'(t),t) \\
\textrm{ where: }Q(x,t) &=& \frac{-\hbar^2 \nabla^2 R}{2m R} \textrm{ is the quantum potential}\\
\label{eq:bohmGR}
\end{eqnarray}

Note that in Bohm's theory the particle velocity is determined by the wavefunction through $ \dot{x} = \nabla  S$, (see for example \cite{holland}) it is not an independent variable.

Linking General relativity to Bohm's interpretation of quantum theory can work because Bohm's theory is expressed in terms of an underlying realist theory. That of course means it has to be non-local, which it is through the non-local dependence of $\psi$ on measurement conditions: boundary conditions that may include the geometry of future measurements. While this sits uneasily in classical physics and even the usual view of general relativity, it is conceptually compatible with general relativity as a 4D theory of spacetime rather than as evolving 3D spaces. It is consistent with attempts to explain quantum theory in terms of spacetime with non-trivial causal structure \cite{hadley97}. Given the measurement conditions (the context) $G_{ab}$ changes smoothly as required.

\section{Conclusion}
Using a quantum wavefunction as a source for a classical theory is physically untenable because it amounts to giving real physical significance to a probability function. A blind mathematical approach that treats the wavefunction as a real classical wave can only give equations that are devoid of physical meaning.

Bohm's interpretation of quantum theory offers a plausible way to combine quantum theory with classical general relativity.

\bibliographystyle{unsrt}
\end{document}